\documentclass[letterpaper,12pt]{article}
\pdfoutput=1
\usepackage[letterpaper,margin=1in]{geometry}
\usepackage{graphicx}
\usepackage{amssymb}
\usepackage{amsmath}
\usepackage{siunitx}
\usepackage[colorlinks=true,
            urlcolor=blue,
            linkcolor=blue,
            citecolor=blue]{hyperref}
\usepackage{lineno}
\usepackage[T1]{fontenc}
\usepackage{cite}
\newcommand{\lambdabar}{{\mkern0.75mu\mathchar '26\mkern -9.75mu\lambda}}
\begin{document} 

\begin{center}
\Large\textbf{%
Dust collapse and bounce in spherically symmetric quantum-inspired gravity models
}
\end{center}

\centerline{Douglas M. Gingrich}

\begin{center}
\textit{%
Department of Physics, University of Alberta, Edmonton, AB T6G 2E1 Canada\\
\smallskip
TRIUMF, Vancouver, BC V6T 2A3 Canada
}
\end{center}

\begin{center}
e-mail:
\href{mailto:gingrich@ualberta.ca}{gingrich@ualberta.ca}
\end{center}

\centerline{\today}

\begin{abstract}
\noindent
We develop an algebraic equation to describe the collapse and possible bounce
of dust in quantum-inspired gravity models with spherical symmetry from
knowledge of the vacuum solution.
Starting from a wide class of spherically symmetric spacetimes, we write down
the covariant Hamiltonian constraints that under dynamical flow give rise to
metrics of many spherically symmetric gravity models.
The constraint equations are solved for the Hamiltonian evolution and simple
equations for the location of the outer boundary of the dust versus time 
and the apparent horizons in terms of metric shape functions are obtained.
The dust density is not assumed to be homogeneous inside the collapsing ball.
Using the developed algebraic equations, we examine several quantum-inspired
gravity metrics to obtain bounce results either previously obtained by
different methods or new results. 
\end{abstract}

\section{Introduction}

In general relativity, pioneering work on gravitational collapse was performed
by Oppenheimer and Snyder~\cite{PhysRev.56.455}. 
The model describes the gravitational collapse of a homogeneous pressureless
star. 
The boundary of the star starts at rest at past infinity and is in free fall
during collapse.
A gravitational singularity is dynamically generated.
However, avoiding singularities is one of the motivations for quantum gravity. 
Besides resolving the classical singularity problem, the quantum theory should
also address dynamical effects such as the in-falling collapse of matter to form 
black holes and trapped surfaces.

In spherically symmetric spacetime, a matter field is needed to study the
collapse of a star.
A simple but effective choice is a perfect fluid of which we choose dust.
A perfect fluid can be modelled by a collection of particles and if the
particles are non-interacting the perfect fluid is pressureless and called
dust. 
The ball of dust is in free fall along geodesics. 
Inside the ball of dust, the density is usually assumed to be homogeneous.
The exterior of the dust ball is taken to be vacuum.

There are generally two approach to solving the dynamics.
One approach is to solve Einstein's equations with the energy-momentum tensor
for a perfect fluid.
The perfect fluid couples in this simple and intuitive way only in the
conventional geometric approach of general relativity.
Another approach is the Hamiltonian formalism in which the dust Hamiltonian
density is taken as the energy density.
These two methods offer different advantages and disadvantages in physical
insight and ease of calculation but give the same results.
However, in those case in which the postulated spacetime vacuum is not a
solution to Einstein's equation, one must resort to the Hamiltonian formalism
to avoid the requirements of exotic matter.

The usual classical approach is to consider the inside of the star surrounded by
a spherical surface to be the Friedman-Lema{\^i}tre-Robertson-Walker (FLRW)
spacetime and match it to the static vacuum solution outside the surface.  
One obtains the Friedmann equation for a scale factor that determines the size
of the star as a function of time.
In the classical case, the homogeneous dust interior can be matched to a
vacuum Schwarzschild exterior.
In the Hamiltonian formalism, usually spherically symmetric spacetime is
minimally coupled to dust and is known as a Lema\^{i}tre-Tolman-Bondi (LTB)
model~\cite{Lemaitre,doi:10.1073/pnas.20.3.169,10.1093/mnras/107.5-6.410}. 

Many approaches using modified LTB models have been applied to quantum-inspired 
gravity
models~\cite{Vaz:2000zb,Kiefer:2005tw,Lasky,Bojowald:2024ium,Bojowald:2009ih,
Giesel:2009jp}.
Particularly, those motivated by loop quantum gravity use the
Ashtekar-Pawlowski-Singh~\cite{PhysRevLett.96.141301} metric in the
interior glued continuously to a static spherically symmetric
spacetime on the exterior. 
The methods often result in the deformed Friedmann equation

\begin{equation}
\left( \frac{\dot{a}}{a} \right)^2 = \frac{8\pi}{3} \rho \left( 1 -
\frac{\rho}{\rho_c} \right), 
\label{eq:FLRW}
\end{equation}
where $a$ is a time dependent scale factor and $\rho$ is the density of the
star.
The Friedmann equation is deformed by an extra $\rho/\rho_c$ term, where
$\rho_c$ is a critical density.  
The critical density term in the Freedman equation induces an effective
pressure and can be related to the time when the gravitational attraction
vanishes. 
The critical density goes to infinity in the classical limit.
If the initial mass is high enough, collapse proceeds as it does classically,
with a trapping horizon forming and enclosing the collapsing matter at some
time. 
The star reaches a maximum density and minimum radius, and then bounces.
The quantum gravity effects stop the collapse, the dust smoothly bounces
back, and there is no singularity.
The geometry continues regularly into an expanding white-hole phase.
It is not our intention here to use~\eqref{eq:FLRW} so we do not critically
review the assumption or validity of the models leading to the deformed
Friedmann equation.

When modifications to general relativity are introduced, the interior
dynamics can often no longer match uniquely to the static
exterior~\cite{Han:2023wxg,Duque:2023syb}. 
This is because the modifications lead to effective pressures.
Moreover, in order to understand horizon formation, we need to impose matching 
conditions. 
A common approach is to couple to a Vaidya-type spacetime.

Another possibility is to use generalized Painlev{\'e}-Gullstrand (PG)
coordinates to describe the metric and the matching conditions will be
manifestly satisfied~\cite{Kanai:2010ae}.
This could allow the full spacetime metric to be written in a single unique
coordinate patch~\cite{Lasky}.
The formation or avoidance of the singularity is independent of the matching to
the exterior.
Generalized PG coordinates are usually a better choice for Hamiltonian
dynamics, particularly for spherically symmetric LTB
spacetime~\cite{Lasky,Duque:2023syb}. 

We use a canonical and covariant framework to study modified spacetimes with
quantum-inspired gravity corrections using generalized PG coordinates.  
The deformed Hamiltonian constraint of spherical gravity can be coupled to
any matter field with a closed hypersurface deformed
algebra~\cite{Bojowald_2010,Bojowald:2023djr}. 
The hypersurface deformed algebra encodes the covariance of the theory.
The structure function transforms properly to define a metric in a covariant
way. 
The model is covariant in that gauge choices in phase space correspond to
coordinate transformations in spacetime. 
The resulting geometry is gauge independent.

A family of vacuum models are presented that ensure the closure of the algebra
and the correct transformation properties of the structure function.
The models being covariant allow the construction of the associated geometry.
The quantum-inspired gravity models can be considered as regularized versions
of the Schwarzschild black hole leading to a transition surface between the
black hole and white hole. 
We do not specifically considering polymeric or holonomy corrections in the
classical Hamiltonian constraint and the resulting modified LTB
spacetimes~\cite{Husain:2021ojz,Lewandowski:2022zce,Husain:2022gwp,Giesel:2022rxi}.  
The gravity corrections we consider are more agnostic.

An outline of this paper is as follows.
Section~\ref{sec:model} describes the action and Hamiltonian constraints
of the effective model.
The dust-time and areal gauges are applied to obtain the physical Hamiltonian. 
The constraint equations are solved in Section~\ref{sec:constraints} leaving
two modified LTB differential equations.
Section~\ref{sec:density} defines the dust density and a Friedmann-like partial
differential equation is obtain for its evolution.
The differential equation is solved for the outer-dust shell resulting in a
simple integral equation that can be solved for a wide class of spherically
symmetric spacetime metrics.
This is the main result of this paper.
Section~\ref{sec:apparent} deals with apparent horizons.
The collapse and possible bounce are calculated in Section~\ref{sec:collapse}
for a few selected metrics.
Section~\ref{sec:summary} offers some concluding remarks.
An appendix~\ref{sec:appendix} is include which calculates the scalar expansion
of geodesics for a general canonical form of the metric. 
We use geometric units of $G = c = 1$ in this paper.
All equations written here reduce to the classical expressions in the
appropriate limit. 
In all plots, a total mass $M = 1$ of dust has been used.

\section{Effective model\label{sec:model}}

In this section, we write down the action and Hamiltonian constraints of the
effective model. 
Coupling to dust provides a natural time variable leading to a physical
Hamiltonian with spatial diffeomorphism symmetry.
In writing the Hamiltonian constraints, we make significant use of the results
of Alonso-Bardaji and Brizuela~\cite{Alonso-Bardaji:2025hda}. 

The procedure leading to our starting point will be described in words here.
The details can be found in~\cite{Alonso-Bardaji:2025hda} but the definition of
the resulting mathematical expressions will follows.
A phase space of fields is defined to covariantly described a spherically
symmetric geometry through its dynamical flow.
The diffeomorphism constraint is the usual classical expression.
The scalar Hamiltonian constraint is written in a general way up to
second-order in spatial derivatives with respect to the radial coordinate.
This requires six undetermined functions of one of the phase space variable.
The Hamiltonian constraint is defined in such a way as to generate the
dynamical flow covariantly and the hypersurface deformed algebra is anomaly
free. 
The six free functions can be picked to generate a geometry of interest.
A similar procedure was initially developed in~\cite{Bojowald:2023xat} leading
to, so called, emergent gravity.
And more recently in~\cite{Zhang:2024khj,Zhang:2024ney}.
The aspect of this formalism that we use here is the inverse problem, or the
reverse-engineered approach.
By specifying a particular seed metric, one can construct a Hamiltonian
constraint that gives by dynamical flow the initial geometry.
By different choices of gauge, the geometry can be expressed in different
charts, although the seed metric was specified using one particular coordinate
choice. 
Up to a canonical transformation, the Hamiltonian constraint is unique.
The above procedure has been performed in~\cite{Alonso-Bardaji:2025hda} for
static and homogeneous charts.
Of particular relevance here is the use of the approach to a generalization of
the Schwarzschild geometry for which we now define the mathematics.

The general diagonal line element for a static spherically symmetric spacetime
can be written in Schwarzschild coordinates as

\begin{equation}
ds^2 = -\left( h_1(x) - \frac{2m}{h_2(x)} \right) dt^2
+ \frac{1}{h_3(x)} \left( h_1(x) - \frac{2m}{h_2(x)} \right)^{-1} dx^2 + x^2
d\Omega^2 ,
\label{eq:metric}
\end{equation}
with shape functions $h_1(x)$, $h_2(x)$, and $h_3(x)$.
By construction, the shape functions are taken to be continuous,
real, and positive over the radial domain of the line element.
For the classical case, the shape functions are $h_1(x) = h_3(x) = 1$ and $h_2
= x$.
For the quantum-inspired theories of interest here, the shape functions are
required to asymptotically approach the classical functions in the large $x$
limit. 
At the minimum surface, $h_3$ and $1/h_2$ should vanish.
The shape functions in the line element have been defined so the ADM mass $m$
will not appear in the Hamiltonian, since $m$ often arises as an integration
constant.
Sometimes to avoid the mass $m$ appearing in the Hamiltonian, we need to
absorber $m$ into another parameter of the model, thus creating a mass-dependent
parameter. 
In principle, we could start with a line element in other coordinate systems
but because of covariance the final results will be independent of this choice.
Since the formalism developed in~\cite{Alonso-Bardaji:2025hda} makes this
choice and since the diagonal line element is very familiar we build on
previous work.  

The corresponding gravity Hamiltonian with dynamical flow leading to this
metric is known~\cite{Alonso-Bardaji:2025hda}.
Furthermore the Hamiltonian has been shown to be covariant and lead to
the correct coordinate charts for different choices of
gauge~\cite{Alonso-Bardaji:2023vtl}.  
See~\cite{Gingrich:2025zwx} for a nontrivial example.
The resulting spacetime has been shown to be the same as that given by
emergent modified gravity~\cite{Bojowald:2023xat}.
Including a dust field, and integrating over $d\Omega$, the action
is~\cite{Husain:2011tk}

\begin{align}
S &= \int dt \int dx \left[ \frac{1}{2} \dot{K}_x E^x + \frac{1}{2}
  \dot{K}_\varphi E^\varphi + 4\pi \dot{T}p_T\right.\nonumber\\
  &\quad \left.\left. - N \left( \mathcal{H}^{(g)} + 
\mathcal{H}^{(d)} \right) - N^x \left( \mathcal{H}_x^{(g)} +
\mathcal{H}_x^{(d)} \right) \right] \right|_{x=\sqrt{E^x}},  
\label{eq:action}
\end{align}
where

\begin{align}
\mathcal{H}^{(g)} &= -\frac{E^\varphi}{2h_2 } (h_1h_2)^\prime -
\frac{K_\varphi^2E^\varphi}{2h_2} (h_2h_3)^\prime + \frac{h_2^2}{2\sqrt{E^x}}
\left( \frac{(E^x)^\prime}{E^\varphi} \right)^\prime\nonumber\\
&\quad - \left( \frac{2h_2}{\sqrt{E^x}} - 3h_2^\prime \right) \frac{h_2}{E^x}
\frac{[(E^x)^\prime]^2}{8E^\varphi} - 2\sqrt{E^x}h_3K_x K_\varphi,\\
\mathcal{H}^{(d)}  &= 4\pi p_T \sqrt{1 + \frac{h_2^2 h_3}{(E^\varphi)^2}
\left(T^\prime\right)^2},\\
\mathcal{H}_x^{(g)} &= -K_x (E^x)^\prime + E^\varphi K_\varphi^\prime,
\label{eq:diff}\\ 
\mathcal{H}_x^{(d)} &= - 4\pi p_T T^\prime.
\label{eq:diffdust}
\end{align}
In all instances $x$ should be written as $x = \sqrt{E^x}$, so that the
constraint is a function of phase-space variables $(E^x, E^\varphi, K_x,
K_\varphi)$ and their derivatives. 
The primes denote differentiation with respect to $x$.
The phase-space variables $E^x(x,t)$ and $E^\varphi(x,t)$ are the 
densitized triad in the radial and angular directions, respectively.
The densitized triads are conjugate to the extrinsic curvature related variables
$K_x(x,t)$ and $K_\varphi(x,t)$.
The dust field is $T(x,t)$ and $p_T(x,t)$ its conjugate momentum.
In general, the dust field contains other contribution but they break the
spherical symmetry and must vanish~\cite{Husain:2011tk,Duque:2023syb}. 
In the action, $N(x,t)$ and $N^x(x,t)$ are the lapse and the radial
component of the shift vector which are Lagrange multipliers and not dynamical. 
Notice that the spacial metric does not appear in the Hamiltonian and gravity
emerges with a spacial metric given by the inverse structure function in the
constraint brackets~\cite{Bojowald:2023xat}.
Equations~\eqref{eq:action}-\eqref{eq:diffdust} define the starting point of
the model. 

Equation~\eqref{eq:action} is the modified gravity coupled to dust.
We have not used Einstein's equations coupled to a perfect fluid, nor a
Hamiltonian form of them, as the modified gravity is not Einstein gravity.
The vacuum solution of the modified gravity does not solve the inhomogeneous
Einstein equations of vacuum.
But the classical limit of the above leads exactly to Einstein gravity.

Since the phase space are fields, the smeared form of the Hamiltonian constraints
are defined as $H_x[s] = \int dx s\mathcal{H}_x$ and $H[s] = \int dx s \mathcal{H}$.
The Hamiltonian constraint satisfies the canonical hypersurface deformation
algebra

\begin{align}
\left\{ H_x[s_1], H_x[s_2] \right\} &= H_x[s_1 s_2^\prime - s_1^\prime s_2],\\
\left\{ H_x[s_1], H[s_2] \right\} &= H[s_1 s_2^\prime],\\
\left\{ H[s_1], H[s_2] \right\} &= H_x[q^{xx}(s_1 s_2^\prime - s_1^\prime s_2)],
\end{align}
with structure function

\begin{equation}
q^{xx} = \frac{h_2^2 h_3}{(E^\varphi)^2}.
\label{eq:sf}
\end{equation}

The line element suitable for canonical analysis is

\begin{equation}
\begin{aligned}
ds^2 &= -N^2 dt^2 + \frac{1}{q^{xx}}(dx + N^x dt)^2 + E^x d\Omega^2\\
&= -\left( N^2 - \frac{(N^x)^2}{q^{xx}} \right) dt^2 + 2 \frac{N^x}{q^{xx}} dx
dt + \frac{1}{q^{xx}} dx^2 + E^x d\Omega^2.  
\label{eq:ADM}
\end{aligned}
\end{equation}
Note that the classical value of the structure function component
$q^{\vartheta\vartheta} = 1/E^x$ has been used. 
The diagonal line element~\eqref{eq:metric} written in general in-falling PG
coordinates is 

\begin{equation}
ds^2 = -\left( h_1 - \frac{2m}{h_2} \right) d\tau^2 - 2 \sqrt{1 - h_1 +
  \frac{2m}{h_2}} dx d\tau + \frac{1}{h_3} dx^2 + E^x d\Omega^2,
\label{eq:PG}
\end{equation}
where $\tau$ is the PG time.
The reference to ``general'' is because the usual flat spatial
slices are modified by a $1/h_3$ coefficient.
The metric coefficients in the static line element~\eqref{eq:PG} can be
compared with the coefficients in the dynamical line element \eqref{eq:ADM}.

We will solve the Hamiltonian equations of motion for all six fields
$q = E^x$, $E^\varphi$, $K_x$, $K_\varphi$, $T$, and $p_T$
using $\dot{q} = \{ q, H[N] + H_x[N^x] \}$, where $q$ is one of the fields and
the dot represents the time derivative. 
The two constraint equations $\mathcal{H}_x = 0$ and $\mathcal{H} = 0$ will
also be used.
First we remove unnecessary degrees of freedom by making two coordinate gauge
choices. 

The dust-time gauge
~\cite{Brown:1994py,Kelly:2020lec,Munch:2021oqn,Husain:2022gwp,Giesel:2022rxi} 
is used to fix the scalar
constraint and the areal gauge to fix the diffeomorphism constraint.
This gauge fixing leaves a physical Hamiltonian with no constraints remaining.
In the dust-time gauge $T = t$ and thus $T^\prime = 0$.
The dust contribution to the diffeomorphism constraint vanishes.
Solving the scalar constraint in the action gives

\begin{equation}
  \mathcal{H}^{(g)} = -4\pi p_T.
  \label{eq:sector}
\end{equation}
Ensuring this gauge condition be preserved by the dynamics gives $N =
1$.
We are thus working in the fluid frame compatible with PG coordinates.
This physically implies that the dust moves along timelike geodesics of the
spacetime.

We next employ the areal gauge $E^x = x^2$.
The diffeomorphism constraint~\eqref{eq:diff} can be solved to give

\begin{equation}
K_x = \frac{E^\varphi K_\varphi^\prime}{2x}.
\label{eq:Kx}
\end{equation}
The equation of motion for $E^x$ is

\begin{equation}
\dot{E}^x = 2 x h_3 K_\varphi + 2x N^x.
\end{equation}
Requiring $E^x = x^2$ be preserved dynamically, $\dot{E}^x = 0$, gives

\begin{equation}
N^x = -h_3 K_\varphi.
\label{eq:shift}
\end{equation}
The shift provides the observer's negative radial velocity.
Since we are working in the fluid (dust) frame, it is the fluid velocity.

Since $E^x$ is time independent, the term $\dot{K}_x E^x$ in the
action~\eqref{eq:action} is a total derivative in time and can be dropped.
Applying all the conditions above gives the gauge fixed action

\begin{equation}
S_\mathrm{GF} = \int dt \int dx \left[ \frac{\dot{K}_\varphi E^\varphi}{2} -
\mathcal{H}_\text{phys} \right],
\end{equation}
where the true physical Hamiltonian is $\mathcal{H}_\text{phys} =
\mathcal{H}^{(g)}$, as there are no constraints remaining. 
There is one physical degree of freedom remaining at each point due to the dust
field. 

\section{Constraint equations and solutions\label{sec:constraints}}

In this section, we write down the two remaining equations of motion and solve
them along with the Hamiltonian scalar constraint to obtain modified LTB
equations. 
The remaining constraint equations before gauge fixing are

\begin{align}
\label{eq:Ephi}
\dot{E}^\varphi &= \frac{N K_\varphi E^\varphi}{h_2} (h_2h_3)^\prime + 2N h_3
\sqrt{E^x} K_x + (N^x E^\varphi)^\prime,\\
\dot{K}_\varphi &= -\frac{N}{2h_2} (h_1h_2)^\prime -\frac{N
K_\varphi^2}{2h_2} (h_2h_3)^\prime + \left( \frac{N h_2^2}{2\sqrt{E^x}}
\right)^\prime \frac{(E^x)^\prime}{(E^\varphi)^2}\nonumber\\
&\quad + \frac{N}{2} \left( \frac{2h_2}{\sqrt{E^x}} - 3h_2^\prime \right)
\frac{h_2}{E^x} \left( \frac{(E^x)^\prime}{2E^\varphi} \right)^2 + N^x
K_\varphi^\prime.  
\end{align}
Replacing $N = 1$, $E^x = x^2$, and using the previous expressions for
$K_x$~\eqref{eq:Kx} and $N^x$~\eqref{eq:shift} gives

\begin{align}
\dot{E}^\varphi &= \left( \frac{(h_2h_3)^\prime}{h_2} - h_3^\prime \right)
K_\varphi E^\varphi - h_3 K_\varphi (E^\varphi)^\prime,\\ 
\dot{K}_\varphi &= -\frac{(h_1h_2)^\prime}{2h_2} - \frac{(h_2h_3)^\prime}{2h_2} 
K_\varphi^2 + \left( \frac{h_2^2}{x} \right)^\prime \frac{x}{(E^\varphi)^2} +
\frac{1}{2} \left( \frac{2h_2}{x} - 3h_2^\prime \right)
\frac{h_2}{(E^\varphi)^2} - h_3 K_\varphi K_\varphi^\prime.  
\end{align}
These are the equations of motion for the LTB family of metrics in the PG gauge.
They hold inside and outside horizons and in the presence or absence of
matter~\cite{Kelly:2020lec}.
This is precisely the sense in which we consider our solution a modified LTB model.

Using the areal gauge and the expression for $K_x$, the physical Hamiltonian
become 
\begin{equation}
\mathcal{H}_\text{phys} = -\frac{(h_1h_2)^\prime}{2h_2} E^\varphi -
\frac{(h_2h_3)^\prime}{2h_2} K_\varphi^2E^\varphi + \frac{h_2^2}{x}
\left( \frac{x}{E^\varphi} \right)^\prime - \frac{1}{2} \left( \frac{2h_2}{x} 
-3h_2^\prime \right) \frac{h_2}{E^\varphi} - h_3 E^\varphi K_\varphi
K_\varphi^\prime. 
\end{equation}

Conservation of $E^\varphi$ in~\eqref{eq:Ephi} gives a solution $E^\varphi = C
h_2$, where the constant $C = 1$ is taken so as not to redefine the mass $m$ in
the metric~\eqref{eq:metric}.
It also allows a closer correspondence between the metric coefficients in
\eqref{eq:ADM} and \eqref{eq:PG}, and also gives the correct classical limit.

With $E^\varphi = h_2$, the Hamiltonian constraint and last equation of motion
become 

\begin{equation}
\mathcal{H}_\text{phys} = -\frac{(h_1h_2)^\prime}{2} - \frac{(h_2h_3)^\prime}{2}
K_\varphi^2 + \frac{h_2^2}{x} \left( \frac{x}{h_2} \right)^\prime
- \frac{1}{2} \left( \frac{2h_2}{x} -3h_2^\prime \right) - h_2 h_3 K_\varphi
K_\varphi^\prime, 
\end{equation}
and
\begin{equation}
\dot{K}_\varphi = -\frac{(h_1h_2)^\prime}{2h_2} - \frac{(h_2h_3)^\prime}{2h_2}
K_\varphi^2 + \left( \frac{h_2^2}{x} \right)^\prime \frac{x}{h_2^2} +
\frac{1}{2} \left( \frac{2h_2}{x} - 3h_2^\prime \right) \frac{1}{h_2} - h_3 
K_\varphi K_\varphi^\prime.  
\end{equation}

We now take a moment to compare our metric~~\eqref{eq:ADM} with the LTB and FRLW
models.
The LTB metric written in diagonal form is

\begin{equation}
ds^2 = -dt^2 + \frac{[R^\prime(x,t)]^2}{1+\varepsilon(r)} dr^2 + R(r,t)^2
d\Omega^2. 
\end{equation}
Introducing the scale factor $a(t)$ and identifying $R(x,t) = a(t) x$ with
$\varepsilon = -kx^2$ gives the FRLW metric.
To better relate the LTB model to the result obtained here, we transform it to 
generalized in-falling PG coordinates using $x = R(r,t)$, to give

\begin{equation}
ds^2 = -dt^2 + \frac{1}{1+\varepsilon(x,t)} (dx + \dot{R} dt)^2 +
x^2 d\Omega^2. 
\end{equation}
Inspection of~\eqref{eq:ADM} and~\eqref{eq:sf} show the models are related by

\begin{equation}
  N^x = -\dot{R} \quad \text{and} \quad
  \sqrt{h_3} h_2 = \sqrt{1 + \varepsilon} E^\varphi.
\end{equation}
This shows again that the shift acts as the negative radial velocity.
The second condition does not satisfy the typical LTB condition $(E^x)^\prime = 
2\sqrt{1 + \varepsilon} E^\varphi$~\cite{Bojowald:2008ja,Bojowald:2009ih}.
However, we may identify $h_3 = 1 + \varepsilon(x,t)$, which can be used to
determine if the metric of interest is a marginal or nonmarginal LTB model.
The homogeneous limit of the nonmarginal models correspond to positive or
negative spatial curvature.
In the classical limit, $h_3 = 1$ and $h_2 = x$ will satisfy the LTB
condition, and hence a LTB spacetime for the marginally bound solution
$\varepsilon = 0$ and the homogeneous assumptions. 

At this point, we specialize to $h_1 = 1$, which still includes a wide class of
metrics.
To reproduce Minkowski space at spatial and null infinity, $h_1 = 1$
is required. 
For $h_1 \ne 0$, a conformal transformation and suitable redefinitions of $h_2$
and $h_3$ can be made to render $h_1$ effectively unity.
We also note that if one is not worried about having the parameter $m$ appearing in
the Hamiltonian, $h_1$ and $h_2$ are not uniquely defined for many metrics.
Simplification with $h_1 = 1$ gives

\begin{equation}
\mathcal{H}_\text{phys} = -\frac{1}{2} (h_2h_3K_\varphi^2)^\prime
\label{eq:hamevo}
\end{equation}
and
\begin{equation}
\dot{K}_\varphi = -\frac{1}{2h_2} (h_2 h_3 K_\varphi^2)^\prime,
\label{eq:Kphievo}
\end{equation}
which indicates

\begin{equation}
\dot{K}_\varphi = \frac{\mathcal{H}_\text{phys}}{h_2}.
\end{equation}
Equation~\eqref{eq:Kphievo} is a first order partial differential equation that
can be solve for $K_\varphi$ given initial and boundary conditions.
Once $K_\varphi$ is known, $N^x$ is determined by~\eqref{eq:shift},
and~\eqref{eq:hamevo} can be solved for $\mathcal{H}_\mathrm{phys}$. 
Solving for $K_\varphi$ gives

\begin{equation}
K_\varphi = \frac{1}{(t-t_0) \sqrt{h_2 h_3}} \int^x dx \sqrt{\frac{h_2}{h_3}} +
C, 
\end{equation}
where $t_0$ and $C$ are constants of integration which we leave unspecified for
now.  

The $x$-component of the shift vector is

\begin{equation}
N^x = - \frac{1}{t-t_0} \sqrt{\frac{h_3}{h_2}} \int^x dx \sqrt{\frac{h_2}{h_3}},
\end{equation}
where we have set $C = 0$ to allow matching later.

The physical Hamiltonian becomes

\begin{equation}
\mathcal{H}_\mathrm{phys} = - \frac{1}{(t-t_0)^2} \sqrt{\frac{h_2}{h_3}} \int^x dx
\sqrt{\frac{h_2}{h_3}}, 
\end{equation}
where any constant in $K_\varphi$ has be removed by the derivative.

With the areal gauge and solution $E^\varphi = h_2$, the structure function now
becomes $q^{xx} = h_3$ and the line element~\eqref{eq:ADM} is

\begin{equation}
ds^2 = -\left( 1 - \frac{(N^x)^2}{h_3} \right) dt^2 + 2
\frac{N^x}{h_3} dx dt + \frac{1}{h_3} dx^2 + x^2 d\Omega^2.   
\end{equation}

\section{Dust density\label{sec:density}}

The energy density\footnote{In general, $\rho$ is a function of
all three spatial coordinates and must be integrated over the angular
coordinates with the square root of the determinant of the spatial metric.}
$\rho$ of the dust field is related to the dust contribution to the scalar
constraint by $p_T = \sqrt{q} \rho$, where $\sqrt{q}$ is the determinant of the
spatial part of the metric.   
The metric components are

\begin{equation}
q_{xx} = \frac{(E^\varphi)^2}{h_2^2 h_3}, \quad
q_{\vartheta\vartheta} = E^x, \quad \text{and} \quad
q_{\varphi\varphi} = E^x \sin^2\vartheta,
\end{equation}
giving the square root of the determinant

\begin{equation}
\sqrt{q} = \frac{E^\varphi E^x}{h_2 \sqrt{h}_3} \sin\vartheta=
\frac{x^2}{\sqrt{h_3}} \sin\vartheta.
\end{equation}
The radial part of the energy density is

\begin{equation}
\rho(x,t) = \frac{h_2\sqrt{h_3} p_T }{4\pi E^x E^\varphi} = -
\frac{\sqrt{h_3}\mathcal{H}_\text{phys}}{4\pi x^2} = \frac{\sqrt{h_2}}{4\pi
(t-t_0)^2  x^2} \int^x dx \sqrt{\frac{h_2}{h_3}}. 
\label{eq:density}
\end{equation}
Notice the dust density has been defined in terms of the dust's
contribution to the scalar constraint divided  by the determinate of the
spatial part of the metric.
By \eqref{eq:sector} the dust scalar constraint is related to the
gravitational constraint and thus to the physical Hamiltonian.
The final expression for the dust density depends on only the shape functions,
and radial coordinate.
The only remaining implicit dependence on the dust is through the time being
the dust field; the dust is acting as the clock.
We expect $\rho(x,t)$ to be positive and monotonic.

Further insight can be obtained by changing the order of the solution of
$\rho(x,t)$.
The energy density\footnote{The energy density $\rho$ without arguments is now
meant to be the radial part only.} follows the Hamiltonian
profile~\eqref{eq:hamevo}: 

\begin{equation}
\rho = \frac{\sqrt{h_3}}{8\pi x^2} (h_2h_3K_\varphi^2)^\prime.
\label{eq:den1}
\end{equation}
In addition,~\eqref{eq:Kphievo} gives

\begin{equation}
\dot{K}_\varphi = -4\pi\rho \frac{x^2}{h_2\sqrt{h_3}}.
\label{eq:Kd}
\end{equation}
We can invert~\eqref{eq:den1} for $K_\varphi$ to get

\begin{equation}
K_\varphi = -\left[ \frac{8\pi}{h_2 h_3} \int^x dx
\frac{x^2}{\sqrt{h_3}}\rho \right]^{1/2} = -\frac{N^x}{h_3}.  
\label{eq:K1}
\end{equation}
The minus sign is chosen so $N^x > 0$ in~\eqref{eq:shift} and the star
collapses in the PG coordinates. 
Choosing the positive root is equivalent to reversing the time coordinate, and
gives an expanding model, or correspondingly, flipping the sign in the PG
coordinate definition.
PG coordinates do not allow for a single branch to be used globally.

The shape function $h_3$ is continuous, real, and positive over the domain.
The shift is continuous, real, and positive over the ingoing branch.
The intrinsic curvature $K_\varphi$ is thus negative over the ingoing branch
except at the minimum surface where it diverges.
This divergence is known to sometime occur in some quantum-inspired models.
Durque~\cite{Duque:2023syb} considers it a maximum-curvature and shows it is a
coordinate singularity of the PG coordinates.
It is not a geometric singularity, but a minimum radius placed at the
maximum-curvature surface.
All the canonical variables are finite on the maximum-curvature surface such as the
structure function, lapse, and shift.
Since we, use PG coordinates we consider the curvature to have a coordinate
singularity at the minimum surface.  

Differentiating~\eqref{eq:K1} with respect to time gives

\begin{equation}
\dot{K}_\varphi = -\sqrt{\frac{2\pi}{h_2 h_3}} \int^x dx \frac{\dot{\rho}
x^2}{\sqrt{h_3}} \left[ \int^x dx \frac{\rho x^2}{\sqrt{h_3}} \right]^{-1/2}.
\end{equation}
Equating this with~\eqref{eq:Kd} leads to

\begin{equation}
\left( \int^x dx\frac{\dot{\rho} x^2}{\sqrt{h_3}} \right)^2 = 8\pi \frac{\rho^2
x^4}{h_2} \int^x dx \frac{\rho x^2}{\sqrt{h_3}}.
\label{eq:pde}
\end{equation}
This is the integral equation to be solved for $\rho(x,t)$ and has
solution~\eqref{eq:density}.
It is a modified version of the deformed Freedman equation~\eqref{eq:FLRW}
that results in many other approaches. 
Our result is missing the critical density term but instead includes an
$x$-dependent factor.  
If $\rho(x,t) = \rho(t)$ was assumed, the density would still have an
$x$-dependent factor in it, thus falsifying the assumption.
The $x$-dependent function only vanishes for $h_3 = 1$ and $h_2 = x$, as in the
classical case and the Freeman equation is obtained.
Only under these conditions will $\rho$ be homogeneous solution to~\eqref{eq:pde}.

The polymer quantization spacetimes motivated by loop quantum gravity do not
have this issue~\cite{Kelly:2020lec}.
The $\rho(x,t) = \rho(t)$ assumption is made.
This allows $\rho$ to be removed from the integral and the $x$ dependence after
integration vanishes in the solution, thus justifying the assumption. 
The polymer corrections introduce trigonometric functions in the LTB equations.
When solving these equations for the density, a differentiation of an inverse
sine function is take which results in an extra term corresponding to the
critical density term in~\eqref{eq:FLRW}. 
The critical density term in the Freeman equation is a result of introducing
quantum corrections using a bounded (trigonometric) form. 

The polymar quantization approach often starts from a noncovariant Hamiltonian
and imposes the PG gauge~\cite{Kelly:2020uwj}.
If instead the Schwarzschild gauge is imposed, the Schwarzschild diagonal line
element results (consistent with $h_3 = 1$ and $h_2 = x$), which can not be
related to the polymar quantization PG line element by a coordinate
transformation. 
This explains the apparent inconsistency between the results.
In Section~\ref{sec:collapse}, our result will be shown to be equivalent to the
polymer quantization result~\cite{Lewandowski:2022zce}. 

\section{Dust evolution\label{sec:evolution}}

In the Oppenheimer-Snyder model of black hole collapse, the space time belongs
to the family of LTB  solutions.
This corresponds to a star of radius $L(t)$ with vacuum outside: $\rho(x,t) =
0$ for $x>L(t)$.
We define $L(t)$ as the comoving radius of the outer dust.
The star is composed of dust within the interior: $\rho(x,t)$ for $x\le L(t)$.  
In specific cases, $\rho(x,t) = \rho(t)$ is a spatially homogeneous density,
but we will need to be more general.
Inhomogeneous dust collapse has been considered in~\cite{Liu:2014kra}.
The inhomogeneities affect the structure of the bounce curve and the trapped
regions.

To describe the vacuum region, we let the energy density vanish at some radius
on the initial hypersurface.
The solution must be diffeomorphic to the Schwarzschild-like metric.
The initial state of the system can be considered as purely classical and all
the quantum corrections can be neglected at the initial time.

For $x > L(t), \rho(x,t) = 0$ and~\eqref{eq:den1} gives $(h_2 h_3
K_\varphi^2)^\prime = 0$.
The only matter constant available is the total mass $M$ of the dust ball. 
We take $2M$ as the constant to allow consistent mass definitions.
Thus $h_2 h_3 K_\varphi^2= 2M$
giving

\begin{equation}
(N^x)^2 =  h_3^2 K_\varphi^2 = \frac{2M h_3}{h_2}.
\label{eq:Nxsta}
\end{equation}
This is consistent with the general PG line element for vacuum~\eqref{eq:PG}
with the identification $h_1 = 1$ when compared to the canonical
form~\eqref{eq:ADM}. 
Then by \eqref{eq:K1} we have matching of the extrinsic curvature
at the boundary.

We can define the effective mass function 

\begin{equation}
  m(x,t) = 4\pi \int^x dx \frac{x^2}{\sqrt{h_3}} \rho,
  \label{eq:mass}
\end{equation}
where $m(x,t)$ is the enclosed mass within a dust shell located at $x$ at time
$t$. 
Once the initial energy density is specified, one can evaluate the initial
mass function through this definition.
This is the mass function commonly defined in LTB spacetimes.
Combined with~\eqref{eq:pde} which is an integral in $x$ and first order in
$t$, we can obtain

\begin{equation}
\dot{m}^2 - \sqrt{\frac{2m h_3}{h_2}} m^\prime = 0,
\end{equation}
which is a general version of the classical result in~\cite{Lasky}.

Using~\eqref{eq:K1} and the effective mass, for $x < L(t)$, we write

\begin{equation}
  (N^x)^2 = \frac{2m(x,t) h_3}{h_2}.
  \label{eq:Nxdyn}
\end{equation}
At $x = L(t)$, the effective mass is time independent and $m(L,t) = M$.
This demonstrates that $(N^x)^2$ is continuous at the boundary.
The $N^x$ expression in~\eqref{eq:Nxdyn} is given by the choice of sign
in~\eqref{eq:shift}, while the $N^x$ expression in~\eqref{eq:Nxsta} is given by
the choice of sign of PG coordinates.
Note that $N^x = 0$ when $h_3 = 0$, that is, the velocity is zero.

The Israel-Darmois junction conditions~\cite{Darmois1927,Israel} must be satisfied at the
outer dust boundary.
Since we are using a pressureless dust, the spatial components, and mixed
components, of the stress-energy tensor vanishes by construction.
The induced metric is continuous across the boundary hypersurface since the
shape functions and the spacial components of the shift vector are continuous at
the boundary.
The extrinsic curvature components are also continuous across the boundary
hypersurface since they also depend on continuity of shape functions and shift
vector. 
Thus there is no non-zero surface stress energy at the boundary.
In addition, the dust field and its normal derivative is continuous at the dust
boundary.

However, for the case of a perfect fluid, not considered here, in which the
surface stress energy is not zero,  the extrinsic curvature would be
discontinuous.
The  pressure is not zero and represents a surface tension.
The surface tension of the fluid sphere balances the internal pressure of the
fluid to keep it stable~\cite{Adler:2005vn,Chakrabarti:2023axz}.

The different layers of dust are labelled by $x$ in the
density~\eqref{eq:density}.
The effective mass using~\eqref{eq:mass} can be calculated as

\begin{equation}
m(x,t) = \frac{1}{2(t-t_0)^2} \left( \int^x dx \sqrt{\frac{h_2}{h_3}} \right)^2.
\end{equation}

The use of the areal gauge implies that a sphere at radius $x$ has surface area
$4\pi x^2$. 
The constancy of dust mass $M$ reflects energy conservation during the
collapse, as no mass is lost to radiation or external forces in this idealized
scenario. 
This relates $\rho$ to the radius of the outermost shell at $x = L(t)$.
The constant $t_0$ is fixed to allow $L = L_0$ (the minimum radius) when $t =
0$.  
The PG metric has two branches.
For $-\infty < t \le 0$, $L(t)$ is decreasing until $t=0$ when
$L(t)=L_0$, then for $0\le t < \infty$, $L(t)$ is increasing.
Equating the mass at the boundary $m(L,t) = M$ gives

\begin{equation}
(t - t_0)^2 = \frac{1}{2M} \left( \int^{L(t)} dx \sqrt{\frac{h_2}{h_3}}
  \right)^2, 
\label{eq:collapse}
\end{equation}
where the initial state of the system must be taken as purely
classical.
This equation for $L(t)$ is the main result of this paper.

The velocity of the outer shell is given by

\begin{equation}
  v = \pm \sqrt{2M} \sqrt{\frac{h_3}{h_2}}.
  \label{eq:velocity}
\end{equation}

The trajectory \eqref{eq:collapse} and velocity \eqref{eq:velocity} are
required to be well behaved over the domain of validity.
Since the shape functions are continuous, real, and positive, the expressions
are well behaved except for possibly at the minimum surface.
The integrand in \eqref{eq:collapse} diverges at the minimum surface but the
square root makes the integral finite.
The velocity vanishes at $h_3 = 0$, where the minimum surface occurs.

So far we have required the shape functions to be continuous, real, and
positive.
For realistic inspired-gravity models, we also require they asymptotically
approach their classical values are large $x$, and that $1/h_2$ and $h_3$ have a zero
at the minimum surface.
Such functions can be represented by power series in inverse power of $x$:

\begin{equation}
  \frac{1}{h_2} = \frac{1}{r} - \frac{1}{2M} \sum_{n=2}^\infty \frac{a_n}{x^n} \quad\text{and}\quad
  h_3 = 1 - \sum_{n=1}^\infty \frac{a_n}{x^n},
\end{equation}
where $a_n$ are meant to be small positive constants relative to $2M$ that
encode the quantum-gravity effects.
Typically the leading term is sufficient.
The minimum surfaces can then be calculated as

\begin{equation}
  x_\mathrm{min} = \left( \frac{a_n}{2M} \right)^{1/(n-1)}
  \quad\text{or}\quad
  x_\mathrm{min} = a_n^{1/n}.
\end{equation}
where we are considering either non-classical $h_2$ or $h_3$ but not both.
It is up to the model to ensure $x_\mathrm{min}$ is positive.
Notice that $L$ versus $t$ will be continuous for all $L\ne L_0$.
If $L_0 = 0$, $L$ versus $t$ will be discontinuous at $t = 0$.
In all cases of a bounce that we will consider, the dust motion is continuous
at the bounce surface.

\section{Apparent horizons\label{sec:apparent}}

We determine the apparent horizons by considering congruences of null geodesics.
The scalar expansions are

\begin{equation}
\theta_{+} = (\ln E^x)^\prime (N\sqrt{q^{xx}} - N^x) \quad\text{and}\quad
\theta_{-} = -(\ln E^x)^\prime (N\sqrt{q^{xx}} + N^x).
\end{equation}
Details of the derivation of these equations can be found in
Appendix~\ref{sec:appendix}.
A trapped surface occurs where both expansions $\theta_\pm$ are negative, and
the boundary of the total trapped region is the apparent horizon.


On the ingoing branch, $N = 1$, $q^{xx} = h_3$, and $N^x = \sqrt{2m(x,t)
  h_3/h_2}$, giving

\begin{equation}
\theta_\pm = \pm \frac{2}{x} \sqrt{h_3} \left( 1 \mp \sqrt{\frac{2m(x,t)}{h_2}}
\right). 
\end{equation}
We see that $\theta_- \le 0$, with equality occurring when $h_3 = 0$.
Likewise $\theta_+ < 0$ for $2m(x,t) > h_2$, and $\theta_+ = 0$ for $2m(x,t) = h_2$
or $h_3 = 0$.
Thus $h_3 = 0$ gives the location of a marginally trapped surface.
It is static and independent of the dust mass.
The other conditions will be examined for some example metrics in the next
section.

On the outgoing branch $N^x = -\sqrt{2(x,t)h_3/h_3}$ and the conditions for an
anti-trapped surface is that both expansions $\theta_\pm$ are positive.
We observe that $\theta_\pm(\text{outgoing branch}) = - \theta
_\mp(\text{outgoing branch})$, so if a trapped surface forms on the incoming
branch, a corresponding anti-trapped surface will form on the outgoing branch. 

\section{Collapse and bounce\label{sec:collapse}}

We now consider some common static and spherically symmetric metrics, and
calculate the dust shell evolution using~\eqref{eq:collapse}.
We also comment on the apparent horizons.
First consider some cases of $h_3 = 1$, which correspond to marginal LTB models. 

\subsection{Schwarzschild spacetime}

The Schwarzschild line element in diagonal form is

\begin{equation}
ds^2 = -\left( 1 - \frac{2M}{r} \right) dt^2 +
\left( 1 - \frac{2M}{r} \right)^{-1} dr^2 + r^2 d\Omega^2,
\end{equation}
where $0 < r < \infty$.
The shape functions from the metric coefficients are $h_1 = 1$, $h_2 = x$, and
$h_3 = 1$. 
The evolution equation~\eqref{eq:collapse} can be solved to give

\begin{equation}
t^2 = \frac{1}{2M} \left( \int^L dx \sqrt{x} \right)^2 = \frac{2}{9M} L^3. 
\end{equation}
A singularity exits at $r = 0$, where the evolution $L(t)$ is discontinuous and
$v\to \pm\infty$.
An apparent horizon forms when $L(t) = 2M$, and a trapped region occurs at $0 <
r < 2M$. 

\subsection{\texorpdfstring{$\bar{\mu}$-loop black hole}{Loop black hole}}

Kelly, Santacruz and Wilson-Ewing~\cite{Kelly:2020uwj} have obtained an
effective loop quantum gravity spacetime using the $\bar{\mu}$-scheme (scale
dependent) holonomy corrections in the areal gauge in PG coordinates.
Note that the model as originally presented is not covariant and different
gauge choices lead to different geometries, which cannot be related to
coordinate transformations~\cite{Alonso-Bardaji:2025hda}. 
The metric in the Schwarzschild gauge is the classical Schwarzschild metric, so 
instead we use the line element in~\cite{Lewandowski:2022zce}:

\begin{equation}
ds^2 = -\left( 1 - \frac{2M}{r} + \frac{4M^2\gamma^2\Delta}{r^4} \right) dt^2 +
\left( 1 - \frac{2M}{r} + \frac{4M^2\gamma^2\Delta}{r^4} \right)^{-1} dr^2 +
r^2 d\Omega^2, 
\end{equation}
where $(2M\gamma^2\Delta)^{1/3} < r < \infty$. 
This metric arises uniquely by matching the Ashtekar-Pawlowski-Singh dust
spacetime to an external spherically symmetric vacuum spacetime outside the
dust~\cite{Lewandowski:2022zce}.

The shape functions from the metric coefficients are $h_1 = 1$, $h_3 = 1$, and

\begin{equation}
h_2 = \frac{x^4}{x^3 - 2\gamma^2\tilde{\Delta}},
\end{equation}
with $\tilde{\Delta} = M\Delta$ giving from~\eqref{eq:collapse}

\begin{equation}
t^2 = \frac{1}{2M} \left( \int^L dx \frac{x^2}{\sqrt{x^3 -
    2\gamma^2\tilde{\Delta}}} \right)^2 
= \frac{2}{9M} L^3 \left(1 - \frac{2\gamma^2\tilde{\Delta}}{L^3} \right).
\end{equation}
Notice the time evolution involves two terms: the classical term and minimum
radius term at which a bounce occurs.
The correct classical limit is obtained when $\Delta\to 0$ and the minimum
radius approaches zero.

This is the same result obtained in~\cite{Kelly:2020lec} by other means.
In that work, $h_2 = x$ (homogeneous dust), is assumed as in the classical case.
The LTB equations are then modified by holonomy corrections inspired by loop
quantum gravity.
The calculation required the derivative of an inverse sine function which gives 
rise to two terms of which one is the additional term in the distorted FLRW
equation~\eqref{eq:FLRW}. 
Starting from the metric arising from the holonomy corrections, we solve the
modified Hamiltonian constrain equations (not classical) to get the same
result.
In our case, the second term that modifies the classical collapse and thus
gives a bounce is a result of the shape function $h_2$ rather than the result
of a bounded function~\cite{Husain:2021ojz}. 

Two apparent horizon $r_\pm$ form when $h_2(r_\pm) = 2m(r_\pm,t)$, or at the
real roots of

\begin{equation}
  r_\pm^4 - 2Mr_\pm^3 + 4\gamma^2\Delta M^2 = 0
  \label{eq:kswapp}
\end{equation}
in agreement with~\cite{Kelly:2020uwj}.
Two roots of \eqref{eq:kswapp} are always complex.
For the case of two real roots, there are two apparent horizons and a trapped
region between them, which becomes anti-trapped on the outgoing branch.
In the limiting case of one real root, $\theta_+ = 0$ forming a marginally
trapped surface.
No real roots and thus apparent horizons are also possible.
The region $x_\mathrm{min} \le r < r_-$ is not trapped.  

Next we consider the case when the line element is not of the $g_{xx} =
1/g_{tt}$ form, or $h_2 = x$, but $h_3\ne 1$.
These are nonmarginal LTB models.

\subsection{Simpson-Visser spacetime}

The Simpson-Visser metric shares many features with quantum-inspired metrics
and is well studied~\cite{Simpson:2018tsi}.
The line element can be written as

\begin{equation}
ds^2 = -\left( 1 - \frac{2M}{r} \right) dt^2 + \left( 1 - \frac{a^2}{r^2}
\right)^{-1} \left( 1 - \frac{2M}{r} \right)^{-1} dr^2 + r^2 d\Omega^2,
\end{equation}
where $a$ is a nonzero parameter and $a < r < \infty$.

The shape functions from the metric coefficients are $h_1 = 1$, $h_2 = x$, and

\begin{equation}
h_3 = \frac{r^2 - a^2}{r^2}
\end{equation}
corresponding to negative $\varepsilon$ (bounce is possible).
Equation~\eqref{eq:collapse} gives

\begin{equation}
t^2 = \frac{1}{2M} \left(\int^L dx \frac{x^{3/2}}{\sqrt{x^2-a^2}} \right)^2.
\end{equation}
We solve this trajectory numerically in Section~\ref{sec:numerical}.

A marginally trapped surface occurs at $r = a$ and an apparent horizon forms 
when $L(t) = 2M$.
The region is trapped on the inward branch and anti-trapped on the outward
branch. 

\subsection{Covariant loop black hole}

In emergent gravity, the line element in the Schwarzschild gauge reads 

\begin{equation}
ds^2 = -\left( 1 - \frac{2M}{x} \right) dt^2 + \left[ 1 + \lambda^2 \left( 1 -
\frac{2M}{x} \right)\right]^{-1} \left( 1 - \frac{2M}{x} \right)^{-1}
\frac{dx^2}{\chi^2} + x^2 d\Omega^2,
\end{equation}
where $\chi$ is an undetermined constant and $\lambda$ is an undetermined
function.
The classical limit is obtained when $\chi\to 1$ and $\lambda\to 0$.

%

\subsubsection{Scale-independent holonomies}

If $\lambda$ is asymptotically constant and we take $\chi^2 = 1/(1+\lambda^2)$, 
the line element becomes

\begin{equation}
ds^2 = -\left( 1 - \frac{2M}{x} \right) dt^2 + \left( 1 - \frac{r_0}{x}
\right)^{-1} \left( 1 - \frac{2M}{x} \right)^{-1} dr^2 + x^2 d\Omega^2 .
\end{equation}
The parameter $r_0$ is defined as

\begin{equation}
r_0 = \frac{2M\lambda}{1+\lambda^2},
\end{equation}
and the holonomy parameter $\lambda$ is scale-independent.

From this line element, $h_1 = 1$, $h_2 = x$, and

\begin{equation}
h_3 = \frac{x - r_0}{x}
\end{equation}
corresponding to negative $\varepsilon$ (bounce is possible).
Equation~\eqref{eq:collapse} gives

\begin{equation}
t^2 = \frac{1}{2M} \left( \int^L dx \frac{x}{\sqrt{x-r_0}} \right)^2 =
\frac{2}{9M} L^3 \left( 1 + \frac{2r_0}{L} \right)^2 \left( 1 - \frac{r_0}{L}
\right). 
\end{equation}
In this case, there is an additional positive function multiplying the
classical result which becomes zero before the classical singularity is reached.
The correct classical limit is obtained when $r_0\to 0$ and the minimum radius
approaches zero.
This is the same result obtained in~\cite{Alonso-Bardaji:2023qgu} by other
means with definitions $r_0 = 2\lambdabar M$ and $\lambdabar =
\lambda^2/(1+\lambda^2)$.   

A marginally trapped surface occurs at $x = r_0$ and an apparent horizon forms 
when $L(t) = 2M$, in agreement with~\cite{Alonso-Bardaji:2023qgu}. 
The region is trapped on the inward branch and anti-trapped on the outward
branch. 

\subsubsection{Scale-dependent holonomies}

Another simple (common) choice is to take $\chi = 1$ and $\lambda^2 = \Delta/x^2$.
The line element is

\begin{equation}
ds^2 = -\left( 1 - \frac{2M}{x} \right) dt^2 + \left[ 1 + \frac{\Delta}{x^2}
\left( 1 - \frac{2M}{x} \right) \right]^{-1} \left( 1 - \frac{2M}{x}
\right)^{-1} dx^2 + x^2 d\Omega^2, 
\end{equation}
where $r_0 < x < \infty$ and $r_0$ is the largest real root of

\begin{equation}
1 + \lambda(r_0)^2\left( 1 - \frac{2M}{r_0} \right) = 0.    
\end{equation}
This line element is the covariant version of the $\bar{\mu}$-loop black hole.

From this line element $h_1 = 1$, $h_2 = x$, and

\begin{equation}
h_3 = \frac{x^3 + \Delta x - 2 \tilde{\Delta}}{x^3}
\end{equation}
where $\tilde{\Delta} = M\Delta$.
In this, case $\varepsilon$ can change sign, going from positive to negative at
$x = 2M$.
Hence a bounce is still possible.
Equation~\eqref{eq:collapse} gives

\begin{equation}
  t^2 = \frac{1}{2M}
  \int^L dx \left( \frac{x^2}{\sqrt{x^3+\Delta x - 2\tilde{\Delta}}}
  \right)^2. 
\end{equation}
We solve this trajectory numerically in Section~\ref{sec:numerical}.
The correct classical limit is obtained when $\Delta\to 0$ and the minimum
radius approaches zero.
This has also been studied in~\cite{Duque:2023syb,Bojowald:2024ium} with the
conclusion of giving a singular geometry.

A marginally trapped surface $x_0$ occurs at the solution of
  
\begin{equation}
x_0^3 + \Delta x_0 -2M\Delta = 0,
\label{eq:covapp}
\end{equation}
and an apparent horizon forms when $L(t) = 2M$, in agreement
with~\cite{Belfaqih:2024vfk}.
Two roots in \eqref{eq:covapp} are complex, leaving only one real root.
The region is trapped on the inward branch and anti-trapped on the outward
branch. 

%

Many more line elements could be considered.

\subsection{Graphical and numerical results\label{sec:numerical}}

In this section, we graphically present the density, mass, and bounce results
for the previously discussed example metrics.

\begin{figure}[p]
\centering
\includegraphics[width=0.49\linewidth]{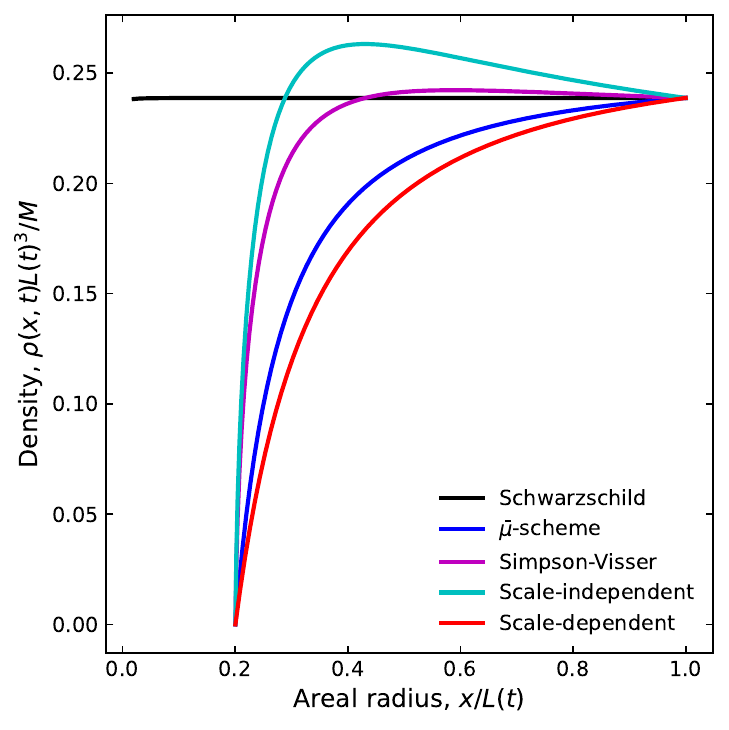}
\includegraphics[width=0.49\linewidth]{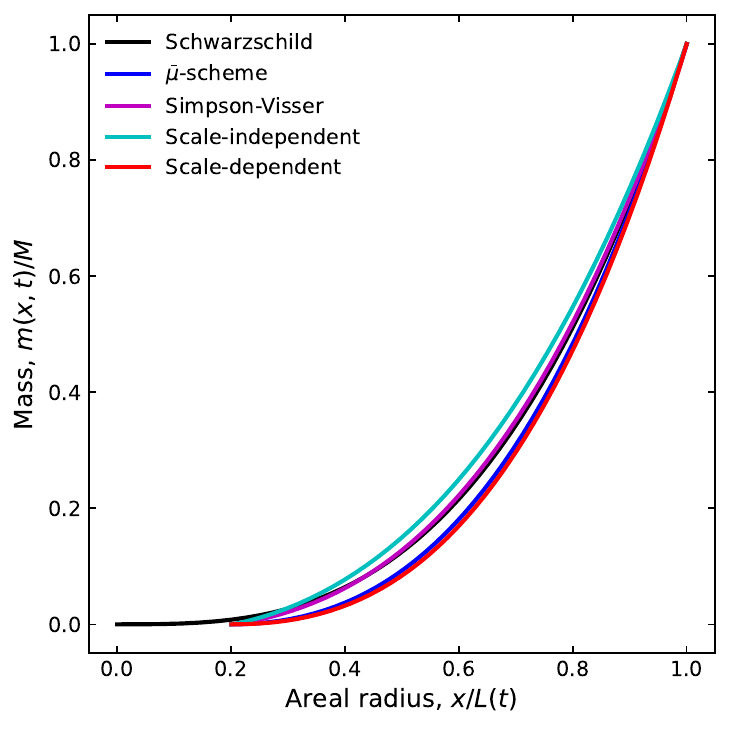}
\caption{Density (left) and mass (right) of dust versus areal radius inside
the outer dust boundary.
The parameters have been chosen to give a common minimum radius (except
Schwarzschild) of $L_0 = 0.2$ for $M = 1$.}     
\label{fig:density}
\end{figure}

Figure~\ref{fig:density}~(left) shows the density versus areal radius inside the
outer shell.
The Schwarzschild density is homogeneous while the quantum-inspired densities
decrease from the dust boundary until they reach zero at the minimum radius.
Figure~\ref{fig:density}~(right) shows the effective mass versus areal radius
inside the outer shell.
All the effecive masses decrease from $M$ at the outer shell to zero at the
minimum radius.

\begin{figure}[p]
\centering
\includegraphics[width=0.49\linewidth]{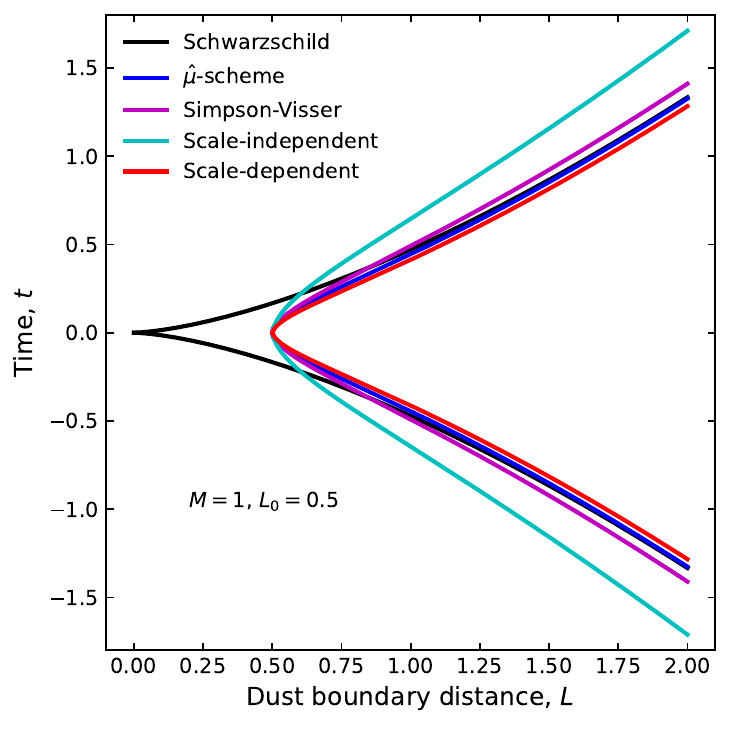}
\includegraphics[width=0.49\linewidth]{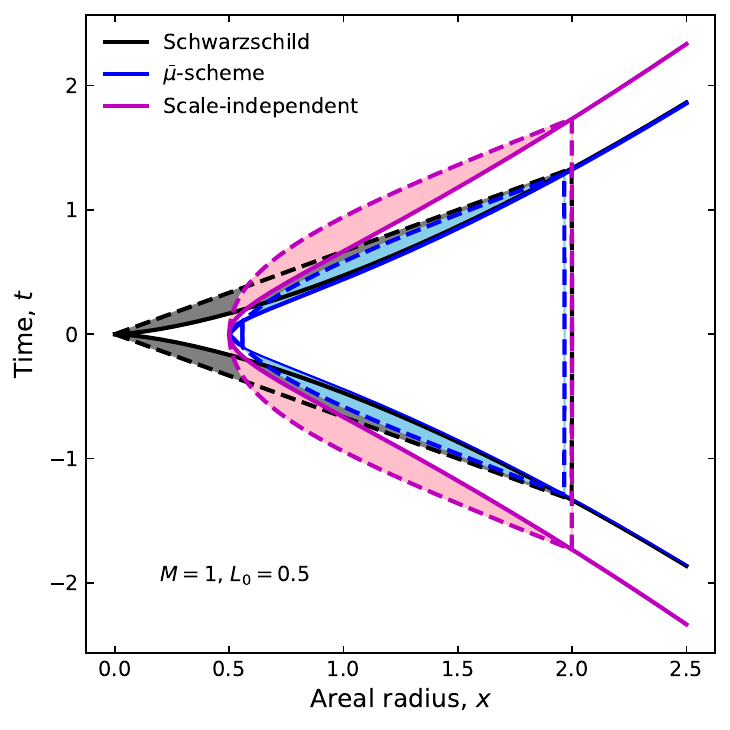}
\caption{(left) Location of the outer boundary of dust $L(t)$ versus time.
  (right) Location of the outer boundary of dust (solid curves) and apparent
  horizons (dashed curves) for three representative metrics.
  The shaded areas are trapped surfaces.
  The parameters have been chosen to give a common minimum radius (except
  Schwarzschild) of $L_0 =   0.5$ with $M = 1$.}    
\label{fig:fig}
\end{figure}

Figure~\ref{fig:fig} (left) shows the collapse of the outer shell of dust for the
example metrics.
The parameters for each quantum-inspired model have been chosen to give a
minimum radius of $L_0 = 0.5$.
We see that the Schwarzschild case is discontinuous at $t = 0$.
Figure~\ref{fig:fig} (right) shows the location of the outer boundary of dust
and apparent horizons for three representative metrics.
The shaded areas are trapped surfaces.
We note the two horizons in the $\bar{\mu}$-loop black hole.

\section{Concluding remarks\label{sec:summary}}

We have obtained an algebraic equation capable of describing spherically
symmetric dust collapse and possible bounce from a knowledge of the static
diagonal line element of the vacuum solution.
The apparent horizons can also be obtained in a similar manner.
Below is a discussion of what has been achieved and acknowledgement of
shortcomings of the model.

We have solved the nonclassical canonical system of constraints and equations
of motion to obtain a dynamical dust density and radial shift vector (negative
velocity) in PG coordinates. 
We made no assumptions about the homogeneity of the dust density.
We did not start from a black hole spacetime or assume a preexisting horizon.
The form of the static metric outside the density ball is the same form as the
dynamical metric inside the dust ball.
This requires matching of the radial shift vector and hence extrinsic curvature
at the boundary.
The approach has similarities to~\cite{Kelly:2020uwj} which started from
classical gravity and applied the $\bar{\mu}$-corrections of loop quantum
gravity. 

Painlev{\'e}-Gullstrand coordinates have an advantage of describing the motion
of in-falling particles across the horizon, but they are not without
shortcomings. 
These coordinates can not provide a global coordinate patch for the
quantum-corrected Oppenheimer-Snyder model as they do in the classical case.
They do not provide a maximal analytic extension of Schwarzschild spacetime
and thus cover only part of the extended manifold~\cite{deHaro:2026ioe}.
There are issues when describing Reissner-Nordstr{\"o}m and
Schwarzschild-anti-deSitter spacetimes, for example, by PG
coordinates~\cite{Faraoni:2020ehi}. 
They may also not work well for black holes with Cauchy horizons or some
quantum-inspired black holes with multiple horizons.

It is important that the metric be continuous across the dust boundary. 
Although $(N^x)^2$ is continuous at the boundary, $N^x$ changes sign in the
solution of the interior at $t = 0$ but does not change sign in the exterior
metric. 
The cross term in the static PG line element~\eqref{eq:PG} is always negative
for in-falling observers. 
In order to describe out-falling, a positive sign in the cross term would be
used. 
A discontinuity in a metric coefficients occurs at the bounce.
This discontinuity was tentatively interpreted as a physical discontinuity
of the gravitational field and it was argued that a shock wave must form as a
consequence~\cite{Kelly:2020uwj,Husain:2021ojz,Munch:2021oqn}.
The PG coordinates mean $t$ is the proper time of observers moving at constant
comoving radial coordinate and $x$ is the comoving radial coordinate in the
dust interior. 
The dust bounces out in the same asymptotic region from where it collapsed.
It is possible to interpret this discontinuity as a discontinuity of the clock
field (coordinate effect), evolving in a continuous geometry.
The geometry in this spacetime is continuous, and the discontinuity in the
metric comes only from using the dust field as a relational clock.
No shock wave is formed~\cite{Fazzini:2023scu}.
While using dust may seem nonideal, other types of fields are not without their
shortcomings~\cite{Bojowald:2005qw}.

In the common approach of using a Friedmann interior, it can be open ($k =
-1$), flat ($k = 0$), or closed ($k = 1$).
Only the closed ($k=1$) case is appropriate since a star must start from a
large radius with a non-zero velocity. 
It is unphysical to start from infinity at rest; the initial rate of change of
density would be zero and the initial moment at the maximum expansion.
Closed ($k=1)$ is inconsistent with PG coordinates in which an observer at
infinity free falls from rest.
Classical PG coordinates can only be used for the $k=0$ case (marginal LTB
spacetime).
As we use PG coordinates starting at rest at infinity, we understanding that the
initial behaviour may not be physical.
Subsequent work should address this issue.


We have not considered energy conditions. They are usually violated in some
form for quantum-inspired spacetimes.  Violation of the weak energy condition,
and thus the null energy condition, is anticipated in quantum-inspired
models. Typically the weak energy condition is satisfied at the beginning of
collapse and is violated as the quantum gravity regime is approached. The
energy conditions are averaged classical inequalities.  They are likely to be
violated in the semiclassical quantum regime because of the microscopic
properties of matter. We can neglect the weak energy condition violation if we
re-interpret the exotic matter content on the right of Einstein’s equations
using the modified vacuum as a semiclassical limit coming from an effective
theory of quantum gravity inducing corrections in the strong-gravity regime
[41]. Violation of the energy conditions is a natural way to avoid the
singularity. Violation in the strong gravity regime may lead to a effective
repulsive force that causes a bounce and thus avoidance of the
singularity. Evaluating the energy conditions is best done on a case by case
basis and is beyond the scope of this paper.

Recently a similar presentation of covariant dynamics based
on~\cite{Zhang:2024khj,Zhang:2024ney} has been presented
in~\cite{Zhang:2025ccx}.
The spherically symmetric case of $g_{xx} = 1/g_{tt}$ is considered.
A more general integral formula for the Dirac observable representing the black
hole mass appears as the limit in the integration.
Example metrics are used to evaluate the integral and present the effective
mass in terms of scalar combinations of canonical fields.
No space time trajectories of matter fields are presented.

Finally, we summarize some minor results of the analysis presented here.
\begin{itemize}
\item The critical density obtained in $\bar{\mu}$-scheme holonomy corrections
  is mathematically a result of the bounded (trigonometric) from of the
  corrections. 
\item The velocity of the dust is proportional to $\sqrt{h_3/h_2}$.
\item Some metrics, such as that resulting from scale-dependent holonomies of
  emergent gravity, have marginal and nonmorginal LTB spacetimes in different
  regions. 
\end{itemize}

To conclude, the quantum gravity effects encoded in the shape functions
{\color{cyan} under certain restrictions} generate a non-singular transition
from a black hole to something similar to a white hole (post-bounce solution). 
General relativity appears to be a singular limit of the models.
The next step could be to study the casual structure and energy conditions
using these shape functions. 


\appendix 
\section{Scalar expansion\label{sec:appendix}}

In this appendix, we consider the congruence of null geodesics and
calculate the scalar expansion using the general spherically symmetric
canonical line element.
Due to spherical symmetry, it is sufficient to consider congruences that are
orthogonal to the surface of concentric 2-spheres defined by constant $x$ and
$t$.

The general line element in canonical form is

\begin{equation}
ds^2 = -\left( N^2 -\frac{(N^x)^2}{q^{xx}} \right) dt^2 + 2 \frac{N^x}{q^{xx}} dt
dx + \frac{1}{q^{xx}} dx^2 + E^x d\Omega^2,
\end{equation}
where we can identify the metric coefficients as

\begin{equation}
  g_{00} = \frac{(N^x)^2}{q^{xx}} - N^2,\quad
  g_{01} = g_{10} = \frac{N^x}{q^{xx}},\quad
  g_{11} = \frac{1}{q^{xx}},\quad
  g_{22} = E^x,\quad
  g_{33} = E^x\sin^2\theta.
\end{equation}
The nonzero inverse metric coefficients are

\begin{equation}
  g^{00} = -\frac{1}{N^2},\
  g^{01} = g^{10} = \frac{N^x}{N^2},\
  g^{11} = -\frac{(N^x)^2}{N^2} + q^{xx},\
  g^{22} = \frac{1}{E^x},\
  g^{33} = \frac{1}{E^x\sin^2\theta}.
\end{equation}
And the square root of the negative of the determinant is

\begin{equation}
\sqrt{-g} = N E^x\sin^2\theta.
\end{equation}

The geodesic equation with metric compatibility is

\begin{equation}
\kappa = -g_{\mu\nu} \frac{dx^\mu}{d\lambda} \frac{dx^\mu}{d\lambda} = \left(
N^2 -\frac{(N^x)^2}{q^{xx}} \right) \left( \frac{dt}{d\lambda}\right)^2
-2 \frac{N^x}{q^{xx}} \frac{dt}{d\lambda} \frac{dx}{d\lambda} -
\frac{1}{q^{xx}} \left( \frac{dx}{d\lambda} \right)^2,
\end{equation}
where $\lambda$ is an affine parameter and $\kappa$ is a constant.

For massless particles, $\kappa = 0$ and we take $\lambda\to t$ to give

\begin{equation}
\left( N^2 -\frac{(N^x)^2}{q^{xx}} \right) -2 \frac{N^x}{q^{xx}} \frac{dx}{dt}
- \frac{1}{q^{xx}} \left( \frac{dx}{dt} \right)^2 = 0.
\end{equation}
The solutions is

\begin{equation}
\frac{dx}{dt} = -N^x \pm N\sqrt{q^{xx}}.
\end{equation}

We denote the tangent vector to the outgoing null geodesics by

\begin{equation}
\ell^\mu = \left( 1, -N^x + N\sqrt{q^{xx}}, 0, 0 \right).
\end{equation}
The other linearly independent null vector that is also orthogonal to the
2-sphere is

\begin{equation}
k^\mu = \left( 1, -N^x - N\sqrt{q^{xx}}, 0, 0 \right),
\end{equation}
which is the tangent vector to ingoing null geodesics.
The covariant vectors are

\begin{equation}
\ell_\mu = \left( \frac{N N^x}{\sqrt{q^{xx}}} -N^2, \frac{N}{\sqrt{q^{xx}}}, 0, 0
      \right) \quad\text{and}\quad 
k_\mu = \left( -\frac{N N^x}{\sqrt{q^{xx}}} - N^2, -\frac{N}{\sqrt{q^{xx}}}, 0, 0
      \right).
\end{equation}
The normalization of the two fields is $\ell^\mu k_\mu = -2N^2$.

The hypersurface metric for 2-spheres is given by

\begin{equation}
h_{\mu\nu} = g_{\mu\nu} + \frac{1}{2N^2} \left( \ell_\mu k_\nu + k_\mu \ell_\nu
\right),
\end{equation}
with coefficients

\begin{equation}
  h_{00} = 0,\quad
  h_{01} = h_{10} = 0,\quad
  h_{11} = 0,\quad
  h_{22} = E^x,\quad
  h_{33} = E^x\sin^2\theta.
\end{equation}
The nonzero inverse coefficients are

\begin{equation}
  h^{22} = \frac{1}{E^x} \quad \text{and} \quad
  h^{33} = \frac{1}{E^x\sin^2\theta}.
\end{equation}

To evaluate covariant derivatives we will need the following Christoffel
symbols:

\begin{align}
\Gamma_{22}^0 &= \frac{1}{2} g^{0\rho} ( \partial_2 g_{2\rho} + \partial_2
g_{\rho 2} - \partial_\rho g_{22}) = -\frac{1}{2} g^{01} \partial_1 g_{22} =
-\frac{1}{2} \frac{N^x}{N^2} (E^x)^\prime,\\
\Gamma_{22}^1 &= \frac{1}{2} g^{1\rho} ( \partial_2 g_{2\rho} + \partial_2
g_{\rho 2} - \partial_\rho g_{22}) = -\frac{1}{2} g^{11} \partial_1 g_{22} =
-\frac{1}{2} \left( -\frac{(N^x)^2}{N^2} + q^{xx} \right) (E^x)^\prime. 
\end{align}

The outgoing and ingoing expansions are given by, respectively,

\begin{equation}
\theta_{+} = h^{\mu\nu} \nabla_\mu \ell_\nu \quad\text{and}\quad
\theta_{-} = h^{\mu\nu} \nabla_\mu k_\nu.
\end{equation}
A calculation gives

\begin{align}
  \theta_+ &=  h^{22} \nabla_2 \ell_2 + h^{33} \nabla_{22} \ell_3\nonumber\\
  & = \frac{1}{E^x} (\partial_2 \ell_2 - \Gamma_{22}^\rho \ell_\rho) +
  \frac{1}{E^x\sin^2\theta} (\partial_3 \ell_3 - \Gamma_{33}^\rho
  \ell_\rho)\nonumber\\
  & = -\frac{1}{E^x} (\Gamma_{22}^0 \ell_0 + \Gamma_{22}^1 \ell_1) +
  -\frac{1}{E^x\sin^2\theta} (\Gamma_{33}^0 \ell_0 + \Gamma_{33}^1
  \ell_1)\nonumber\\ 
  & = -\frac{2}{E^x} (\Gamma_{22}^0 \ell_0 + \Gamma_{22}^1 \ell_1).
\end{align}
The final result is

\begin{equation}
\theta_{+} = (\ln E^x)^\prime (N\sqrt{q^{xx}} - N^x) \quad\text{and}\quad
\theta_{-} = -(\ln E^x)^\prime (N\sqrt{q^{xx}} + N^x).
\end{equation}
These general formula reduce to the usual specific forms.

\section*{Acknowledgments}
We acknowledge the support of the Natural Sciences and Engineering
Research Council of Canada (NSERC). 
Nous remercions le Conseil de recherches en sciences naturelles et en
g{\'e}nie du Canada (CRSNG) de son soutien. 
\bibliographystyle{JHEP}
\bibliography{gingrich}
\end{document}